\documentclass[traditabstract,a4,letter]{aa}  

\usepackage{epsfig}
\usepackage{graphicx}
\usepackage{epstopdf}
\usepackage{color}  
\usepackage{array}   
\usepackage{units}  
\usepackage[breaklinks,colorlinks, citecolor=blue]{hyperref}
\usepackage{natbib}  
\usepackage{multirow}   
\usepackage{amsmath,amssymb}  
\usepackage[english]{babel}
\usepackage[latin1]{inputenc}
\usepackage[T1]{fontenc}
\usepackage{longtable,lscape}
\usepackage{footnote}

\usepackage{txfonts}
\usepackage[toc,page]{appendix} 

\begin{document}


\title{Predicting the CIB-$\phi$ contamination in\\ the cross-correlation of the tSZ effect and $\phi$}
\author{G.Hurier\inst{1}}

\institute{
Institut d'Astrophysique Spatiale, CNRS (UMR8617) and Universit\'{e} Paris-Sud 11, B\^{a}timent 121, 91405 Orsay, France 
\\
\email{ghurier@ias.u-psud.fr} 
}

   \abstract{The recent release of {\it Planck} data gives access to a full sky coverage of the thermal Sunyaev-Zel'dovich (tSZ) effect and of the cosmic microwave background (CMB) lensing potential ($\phi$).
   The cross-correlation of these two probes of the large-scale structures in the Universe is a powerful tool for testing cosmological models, especially in the context of the difference between galaxy clusters and CMB for the best-fitting cosmological parameters.\\
   However, the tSZ effect maps are highly contaminated by cosmic infra-red background (CIB) fluctuations.
   Unlike other astrophysical components, the spatial distribution of CIB varies with frequency. 
   Thus it cannot be completely removed from a tSZ Compton parameter map, which is constructed from a linear combination of multiple frequency maps.
  We have estimated the contamination of the CIB-$\phi$ correlation in the tSZ-$\phi$ power-spectrum.
   We considered linear combinations that reconstruct the tSZ Compton parameter from {\it Planck} frequency maps.
   We conclude that even in an optimistic case, the CIB-$\phi$ contamination is significant with respect to the tSZ-$\phi$ signal itself.\\
   Consequently, We stress that tSZ-$\phi$ analyses that are based on Compton parameter maps are highly limited by the bias produced by CIB-$\phi$ contamination.}


\authorrunning{G.Hurier et al.}
\titlerunning{CIB-$\phi$ contamination in\\ the cross-correlation of the tSZ effect and $\phi$}

\maketitle
  
\section{Introduction}

Modern cosmology extensively used cosmic microwave background (CMB) data.
During their propagation along the line of sight, photons are affected by several physical processes such as the thermal Sunyaev-Zel'dovich effect \citep[tSZ,][]{sun69,sun72} and CMB gravitational lensing \citep{bla87}, which trace the gravitational potential integrated along the line of sight, $\phi$.\\
At microwave frequencies, foreground emissions contribute to the total signal of the sky, for example the cosmic infra-red background \citep[CIB,][]{pug96,fix98}.
The tSZ effect, gravitational lensing, and the CIB are tracers of the large-scale structures in the matter distribution of the Universe. 
They have been powerful sources of cosmological and astrophysical constraints \citep[see e.g.,][]{planckSZC,PlanckCIB,PlanckPHI}.\\

All these probes present correlations through their relation to the large-scale distribution of matter in the Universe.
The CIB-$\phi$ correlation has already been detected with high significance \citep{PlanckCIBPHI}.
More recently, the tSZ-$\phi$ correlation estimated from a Compton parameter map \citep{hil13} has been used to constrain cosmological parameters.\\

We discuss the contamination from the CIB-$\phi$ correlation into the tSZ-$\phi$ correlation estimated from a Compton parameter map.
In Sect.~\ref{crossth}, we present a coherent modelling of the tSZ, $\phi$, and the CIB spectra and cross-correlation.
Then in Sect.~\ref{secconta}, we estimate the CIB-$\phi$ contamination in tSZ-$\phi$ cross-correlation.
In Sect.~\ref{secdis}, we discuss the cleaning process proposed by \citet{hil13} and estimate the residual contamination level.
Finally in Sect.~\ref{seccon}, we conclude.\\

Throughout the paper, we consider $H_0 = 67.1$, $\sigma_8 = 0.80$ and $\Omega_{\rm m} = 0.32$ as our fiducial cosmological model, unless otherwise specified.

\section{Modelling the tSZ, $\phi$, and CIB cross-correlations}
\label{crossth}

The total power-spectrum between the  A and B quantities (tSZ, $\phi$ or CIB) can be expressed as
\begin{equation}
C_{\ell}^{\rm A\times B} = C_{\ell}^{\rm A\times B-{P}} + C_{\ell}^{\rm A\times B-{C}},
\end{equation}
with $C_{\ell}^{\rm A\times B-{P}}$ the Poissonian contribution and $C_{\ell}^{\rm A\times B-{C}}$ the contribution from large angular scale correlations.

\subsection{Poissonian term}
\label{1-halo}
The Poissonian or one-halo term of the power-spectrum can be written using a halo model as the sum of the contributions from each halo. 
The number of halos per unit of mass and redshift, $\frac{{\rm d^2N}}{{\rm d}M {\rm d}V}$, is given by the mass function \citep[see e.g.,][]{tin08}.\\
Using the limber approximation, we can write the one-halo term as
\begin{equation}
C_{\ell}^{\rm A\times B-{P}} = 4 \pi \int {\rm d}z \frac{{\rm d}V}{{\rm d}z {\rm d}\Omega}\int{\rm d}M \frac{{\rm d^2N}}{{\rm d}M {\rm d}V} W^{\rm P}_{\rm A} W^{\rm P}_{\rm B}.
\end{equation}

\noindent The tSZ contribution can be written as
\begin{equation}
W^{\rm P}_{\rm tSZ} = Y_{500} y_{\ell},
\end{equation}
with $Y_{500}$ the tSZ flux of the clusters, related to the mass, $M_{500}$ via the scaling law presented in Eq.~\ref{scale}, and $y_\ell$ the Fourier transform on the sphere of the cluster pressure profile from \citet{planckSZC} per unit of tSZ flux.\\
\noindent We used the $M_{500}-Y_{500}$ scaling law presented in \citet{planckSZC}, 
\begin{equation}
E^{-\beta_{\rm sz}}(z) \left[ \frac{D^2_{ang}(z) {Y}_{500}}{10^{-4}\,{\rm Mpc}^2} \right] = Y_\star \left[ \frac{h}{0.7} \right]^{-2+\alpha_{\rm sz}} \left[ \frac{(1-b) M_{500}}{6 \times 10^{14}\,{\rm M_{\odot}}}\right]^{\alpha_{\rm sz}},
\label{scale}
\end{equation}
with $E(z) = \Omega_{\rm m}(1+z)^3 + \Omega_{\Lambda}$. The coefficients $Y_\star$, $\alpha_{\rm sz}$ and $\beta_{\rm sz}$, taken from \citet{planckSZC}, are given in Table~\ref{tabscal}.
The mean bias, $(1-b)$, between X-ray mass and the true mass was estimated from numerical simulations, b $\simeq$ 0.2 \citep[see][for a detailed discussion of this bias]{planckSZC} .\\

\noindent The lensing contribution can be written as 
\begin{equation}
W^{\rm P}_{\phi}~=~-2 \psi_\ell \frac{(\chi' - \chi) \chi}{\chi'},
\end{equation}
with $\chi$ the comoving distance, $\chi'$ the comoving distance of the CMB, and $\psi_\ell$ the 3D lensing potential Fourier transform on the sky.\\
We can express the potential $\psi$ as a function of the density contrast,
\begin{equation}
\Delta \psi = \frac{3}{2} \Omega_m H_0^2 \frac{\delta}{a},
\end{equation}
with $a$ the universe scale factor and $\delta$ the density contrast. Then, the lensing contribution reads
\begin{equation}
W^{\rm P}_{\phi} = \frac{3 \Omega_m H_0^2 (1+z)}{\ell (\ell+1)} \frac{(\chi' - \chi) \chi}{\chi'} \delta_\ell,
\end{equation}
where  $\delta_\ell$ is the Fourier transform of the density contrast profile.\\

\noindent The CIB contribution can be written as
\begin{equation}
W^{\rm P}_{\rm CIB}(\nu) = S_{500}(\nu) c_\ell,
\end{equation}
with ${S}_{500}(\nu) = \frac{a\, L_{500}(\nu)}{4 \pi \chi^2(z)}$, the infra-red flux at frequency $\nu$ of the host halo and $c_\ell$ the Fourier transform of the infra-red profile.\\
To model the $L_{500}-M_{500}$ relation we used a parametric relation derived from the relation proposed in \citet{sha12}.
To compute the tSZ-CIB correlation, we need to relate tSZ halos with CIB emissions. Consequently, we cannot use a parametrization that relates galaxy halos with the CIB flux. We need a relation that relates the galaxy cluster halo to the associated total CIB flux.\\
We chose to parametrize the CIB flux, $L_{500}$, at galaxy cluster scale with a power law of the mass\footnote{We consider the CIB-$\phi$ and tSZ-CIB cross-correlations, consequently there is no need to model of the CIB at galaxy scale.}, $M_{500}$,  
\begin{equation}
L_{500}(\nu) = L_0 \left[ \frac{M_{500}}{1 \times 10^{14}\,{\rm M_{\odot}}} \right]^{\epsilon_{\rm cib}}(1+z)^{\delta_{\rm cib}} \, \Theta \left[ (1+z)\nu,T_{\rm d}(z) \right],
\end{equation}
where $L_0$ is a normalization parameter, $T_{\rm d}(z) = T_0 (1+z)^{\alpha_{\rm cib}}$ is the thermal dust temperature, and $\Theta \left[ \nu,T_{\rm d}\right]$ is the typical spectral energy distribution (SED) of a galaxy that contributes to the total CIB emission, 
$$
\Theta \left[ \nu,T_{\rm d}\right] = \left\{
    \begin{array}{ll}
        \nu^{\beta_{\rm cib}} B_{\nu}(T_{\rm d}) & \mbox{if} \  \nu < \nu_0 \\
        \nu^{-\gamma_{\rm cib}} & \mbox{if} \  \nu \geq \nu_0,
    \end{array}
\right.
$$
with $\nu_0$ the solution of $\frac{{\rm d\, log} \left[\nu^{\beta_{\rm cib}} B_{\nu}(T_{\rm d})\right] }{{\rm d\, log}(\nu)} = -\gamma_{\rm cib}$.
The coefficients $T_0$, $\alpha_{\rm cib}$, $\beta_{\rm cib}$, $\gamma_{\rm cib}$, $\delta_{\rm cib}$, and $\epsilon_{\rm cib}$ are given in Table~\ref{tabscal}.

\begin{savenotes}
\begin{table}
\center
\caption{Cosmological and scaling-law parameters for our fiducial model for both $Y_{500}-M_{500}$ \citep{planckSZC} and the $L_{500}-M_{500}$ relation fitted on CIB spectra.}
\begin{tabular}{|cc|cc|cc|}
\hline
\multicolumn{2}{|c|}{Fiducial cosmo} &\multicolumn{2}{|c|}{$M_{500}-Y_{500}$} & \multicolumn{2}{c|}{$M_{500}-L_{500}$} \\
\hline
$\Omega_{\rm m} $&$0.32$  & ${\rm log}\,Y_\star$ & -0.19 $\pm$ 0.02  & $ T_0$ & 24.4\footnote{\label{f2}Fixed value from \citet{PlanckCIB}}\\
$\sigma_8$& $0.80$ &$\alpha_{\rm sz}$ & 1.79 $\pm$ 0.08 & $\alpha_{\rm cib}$ & $0.36^{\ref{f2}}$ \\
$H_0$& $67$&$\beta_{\rm sz}$ & 0.66 $\pm$ 0.5 & $\beta_{\rm cib}$ & $1.75^{\ref{f2}}$\\
&& & & $\gamma_{\rm cib}$ & $1.7^{\ref{f2}}$\\
&& & & $\delta_{\rm cib}$ & $3.2^{\ref{f2}}$ \\
 && & & $\epsilon_{\rm cib}$ & $1.1 \pm 0.2$\\
\hline
\end{tabular}
\label{tabscal}
\end{table}
\end{savenotes}

\subsection{Large-scale correlation terms}
\label{2halos}
We now focuss on the contribution from large angular scale correlations in the matter distribution to the tSZ-$\phi$-CIB cross-correlation power-spectra matrix. We expres this contribution as
\begin{equation}
C_{\ell}^{\rm A\times B-C} = 4 \pi  \int {\rm d}z \frac{{\rm d}V}{{\rm d}z{\rm d}\Omega} W^{\rm C}_{\rm A} W^{\rm C}_{\rm B} P_k,
\end{equation}
with $P_k$, the matter power-spectrum.\\
In the following subsections we present the window functions, $W$, related to each contribution.\\

For the tSZ effect, we considered the two-halo large-scale correlation. We neglected the contribution produced by diffuse tSZ emission from the warm hot interstellar medium.
The two-halo contribution reads
\begin{equation}
W^{\rm C}_{\rm tSZ} = \int{\rm d}M \frac{{\rm d^2N}}{{\rm d}M {\rm d}V} Y_{500} y_{\ell} b_{\rm lin},
\end{equation}
where $b_{\rm lin}$ is the bias relating the halo distribution to the overdensity distribution. We considered the bias from \citet{mo96} and \citet{kom99}, which is realistic at galaxy cluster scale. \\

For the lensing potential, we accounted for the large-scale correlation by considering structures in the linear growth regime as
\begin{equation}
W^{\rm C}_{\phi} = \frac{3 \Omega_m H_0^2}{c^2 \ell (\ell+1)} (1+z) \frac{(\chi' - \chi)}{\chi' \chi}.
\end{equation}\\

For the CIB at frequency $\nu$, we considered the two-halo large-scale correlation as
\begin{equation}
W^{\rm C}_{\rm CIB}(\nu) = \int{\rm d}M \frac{{\rm d^2N}}{{\rm d}M {\rm d}V} S_{500}(\nu) c_{\ell} b_{\rm lin}.
\end{equation}

It is worth noting that the CIB-$\phi$ cross-correlation is dominated by the two-halo term\footnote{This can be explained by the high-$z$ and low-mass halo sensitivity of this cross-correlation, which favours the two-halo, whereas low-$z$, high-mass objects favour the one-halo term.}, in contrast to the tSZ-CIB and tSZ-$\phi$ correlations which present a significant contribution from both one and two-halo terms.

\subsection{Redshift distribution}
\begin{figure}[!h]
\begin{center}
\includegraphics[scale=0.2]{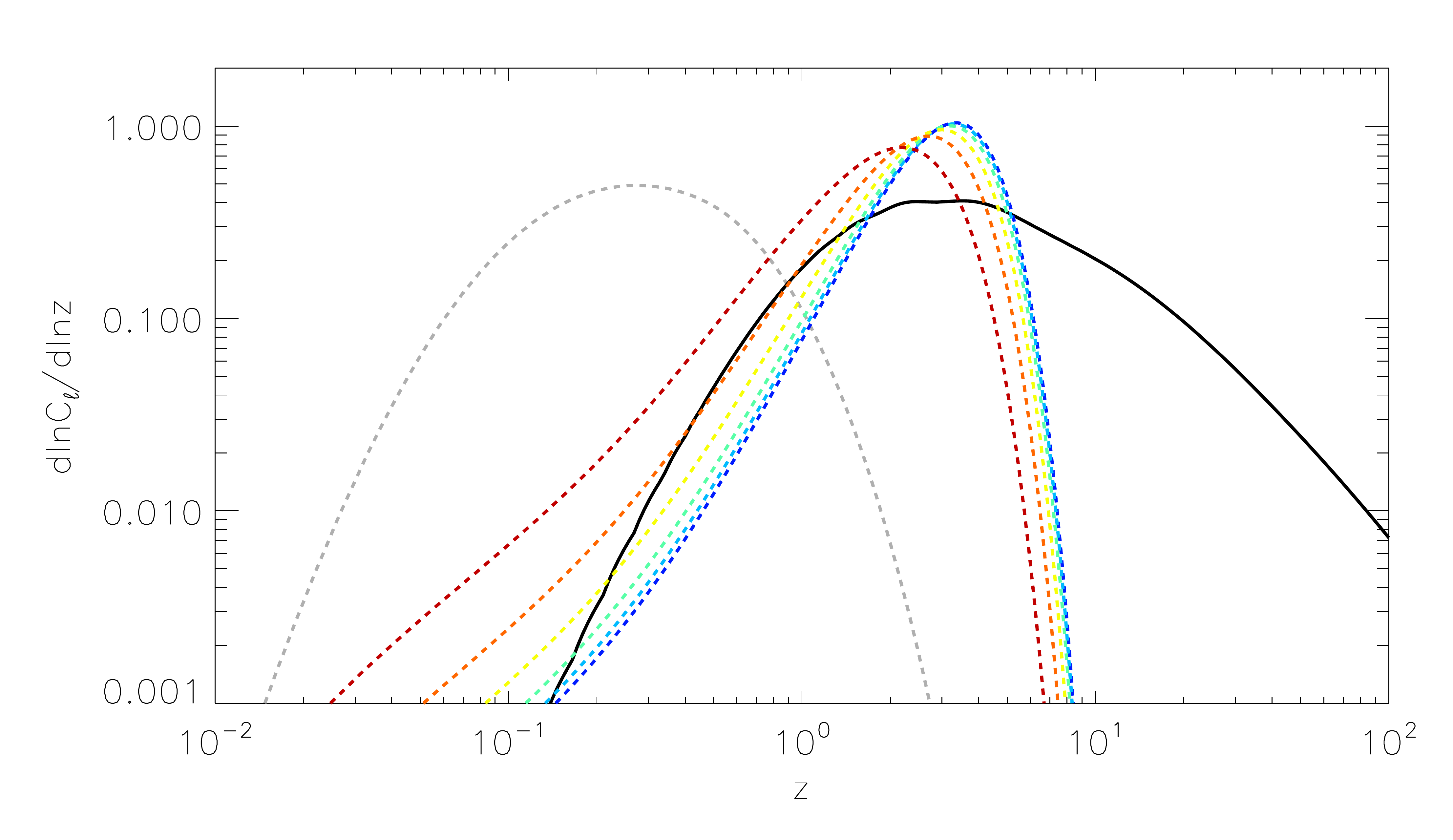}
\caption{Distributions of the tSZ, $\phi$, and CIB power as a function of the redshift at $\ell = 1000$. The black solid line depicts the lensing potential, $\phi$, the grey dashed line the tSZ effect, the dark blue, light blue, green, yellow, orange, and red dashed lines plot the CIB at 100, 143, 217, 353, 545, and 857 GHz, respectively.}
\label{reddis}
\end{center}
\end{figure}
In Fig.~\ref{reddis}, we present the redshift distribution of power for the tSZ, CIB, and lensing potential auto-correlation spectra.
The tSZ effect is produced by structures at low redshift (z < 2). The CIB and $\phi$ spectra are dominated by objects at higher redshift.\\
The mean redshift of structures that dominate in the CIB is a function of the considered frequency.
These redshift windows highlight the high degree of correlation between the CIB and $\phi$ and the low degree of correlation between tSZ and the other two probes of large-scale structures.\\
For example, the high degree of correlation between the CIB and $\phi$ have recently been used to detect the lensing B modes of polarization by the SPT collaboration \citep{han13}.\\
Consequently, small CIB residuals in a tSZ Compton parameter map can lead to a significant bias in the tSZ-$\phi$ correlation analysis.

\section{CIB-$\phi$ contamination in tSZ-$\phi$ cross-correlation}
\label{secconta}
\subsection{tSZ Compton parameter map}

The recently released {\it Planck} data allow constructing full-sky tSZ Compton parameter map \citep{PlanckSZS} using internal linear combination (ILC) component separation methods \citep[see e.g.,][]{rem11,hur13}.
These methods reconstruct the tSZ signals using linear combinations
\begin{equation}
y = \sum_{\nu} w_{\nu} T_{\nu},
\end{equation}
where $T_{\nu}$ is the intensity at the frequency $\nu$ and $w_{\nu}$ are the weights of the linear combination. For ILC-based methods, these weights are computed by minimizing the variance of the reconstructed $y$-map under constraints.\\
The level of CIB-$\phi$ contamination in a tSZ-$\phi$ analysis based on a tSZ Compton parameter map depends on the weights, $w_\nu$, used for the linear combination.\\
We here considered the weights provided in Table 2 of \citet{hil13}, which have been used to constrain the cosmology from the tSZ-$\phi$ cross-correlation power-spectrum.
 
\subsection{Propagation of CIB-$\phi$ spectra}

We stress that the CIB is composed of different populations of objects at different frequencies. The associated power-spectra present variations of shape with respect to the frequency. The CIB-$\phi$ spectra present the same variations of shape with frequency.
The CIB-$\phi$ contamination in the tSZ-$\phi$ cross-correlation reads
\begin{equation}
C^{\rm conta}_\ell = \sum_\nu w_\nu C^{\rm \phi \times CIB(\nu) }_\ell.
\end{equation}

\begin{figure}[!h]
\begin{center}
\includegraphics[scale=0.2]{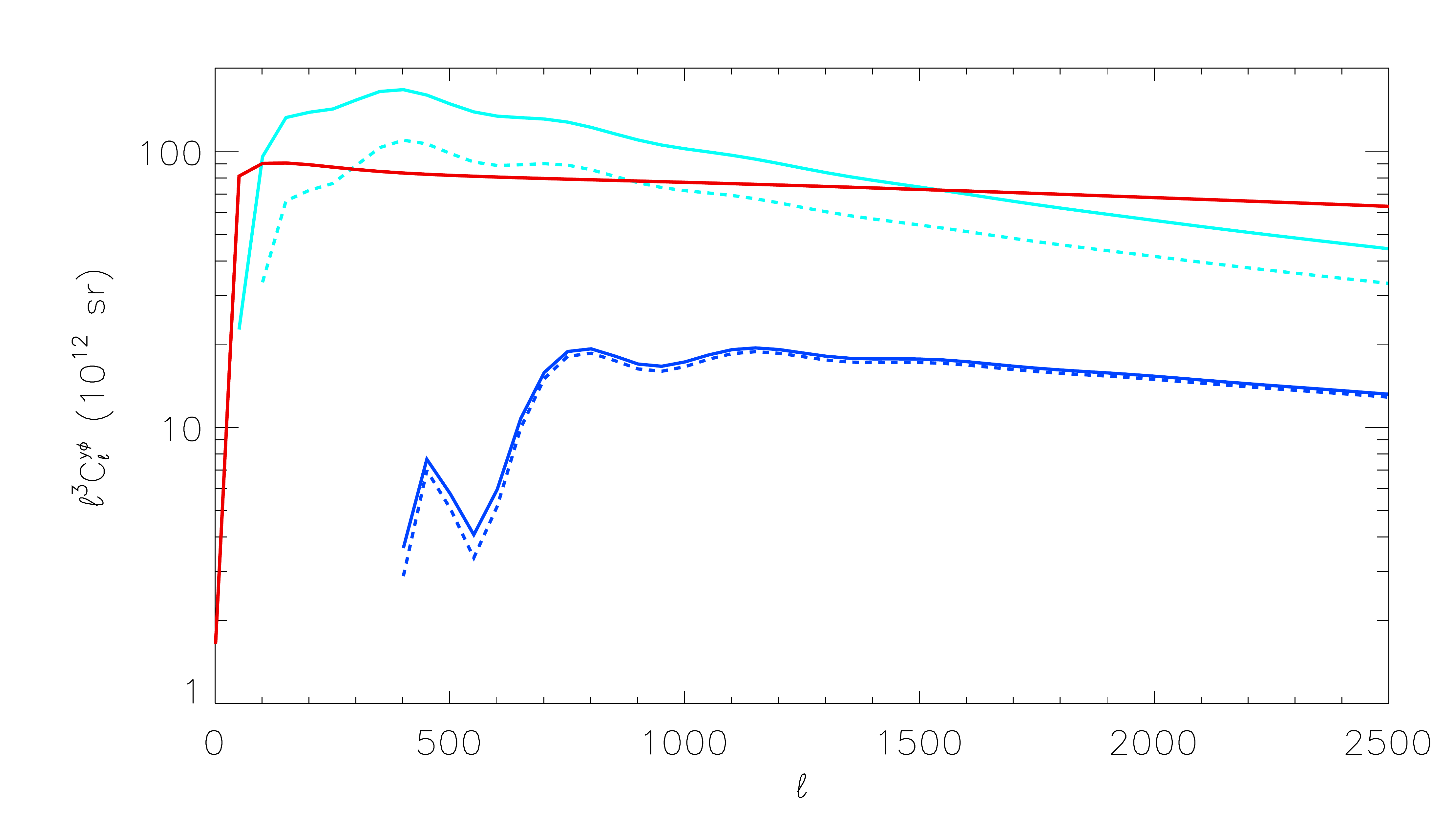}
\caption{Residual contamination before cleaning in light blue and after cleaning in dark blue. The red solid line presents the tSZ-$\phi$ correlation for our fiducial model.  The solid blue lines are the contamination considering the set of weights from \citet{hil13} for $f_{\rm sky} = 0.3$. The dashed blue lines are the contamination considering the set of weights from \citet{hil13} for $f_{\rm sky} = 0.2$.}
\label{cibres}
\end{center}
\end{figure}

In Fig.~\ref{cibres}, we present our estimation of the CIB-$\phi$ contamination considering the set of weights from \citet{hil13} and compare it with the expected tSZ-$\phi$ cross-correlation power-spectrum for our fiducial model.\\
The contamination is high, with a similar amplitude as the tSZ-$\phi$ correlation.
Furthermore, the contamination level is highly dependent on the set of weights. For different sets of weights, the contamination varies an the same order of magnitude as the tSZ-$\phi$ signal.\\
This variability highlights the difficulty of controlling the CIB-$\phi$ contamination that can lead to a high bias in tSZ-$\phi$ analysis.\\
However, the contamination level is highly dependent on the exact CIB frequency correlation matrix. Consequently, the derived value has to be considered as an order-of-magnitude estimate of the effective contamination.

\section{Cleaning the CIB-$\phi$ contamination}
\label{secdis}
In their analysis, \citet{hil13} proposed a method of cleaning for CIB-$\phi$ contamination in tSZ-$\phi$ spectrum.
Their method consists of a correction of the form
\begin{equation}
C_\ell^{\phi\times y,\, {\rm corr}} = C_\ell^{\phi \times y} - \alpha_{\rm corr} C_\ell^{\rm \phi\times CIB(857)},
\end{equation}
where $C_\ell^{\phi \times y}$ is the measured correlation between the tSZ and potential maps. The factor $\alpha_{\rm corr}$ is computed from an adjustment of $C^{\rm CIB(857) \times CIB(857)}_\ell$ on the cross-correlation of the tSZ Compton parameter and the 857 GHz maps.\\
We propose to test the accuracy of this procedure using our modelling of the tSZ, $\phi$, and CIB cross-correlations.
In their analysis \citet{hil13} neglected the tSZ-CIB correlation contribution for computing $\alpha_{\rm corr}$. However, at the concerned frequencies, the tSZ-CIB correlation factor can reach about 20\% and therefore cannot be neglected.
Neglecting tSZ-CIB correlation for the cleaning process leads to an underestimation of the CIB-$\phi$ contamination.
In an optimistic case\footnote{Without uncertainties or systematics for the computation of $\alpha_{\rm corr}$}, $\alpha_{\rm corr}$ reads
\begin{equation}
\alpha_{\rm corr} = \left <\frac{\left(\sum w_\nu C_\ell^{\rm CIB(\nu) \times CIB(857)}\right)+C_\ell^{\rm tSZ \times {CIB}(857)}}{C_\ell^{\rm CIB(857) \times CIB(857)}}\right > .
\end{equation}
We computed this value for each set of weights at $\ell = 600$. 
The value of $\alpha_{\rm corr}$ varies by 3\% depending on the value of $\ell$ used for the computation ($300<\ell$<1500), with a higher level of contamination using higher $\ell$ values.
\citet{hil13} found $\alpha_{\rm corr} = (6.7 \pm 1.1) \times 10^{-6} \,{\rm K}^{-1}_{\rm CMB}$ for $f_{\rm sky} = 0.3$. Considering the associated set of weight we predict $\alpha_{\rm corr} = (7.1 \pm 0.2) \times 10^{-6}\, {\rm K}^{-1}_{\rm CMB}$ for $300 < \ell < 1500$.
With this, we estimate the associated residual contamination in the tSZ-$\phi$ spectrum as
\begin{equation}
C_\ell^{\rm res} = \left( \sum_\nu w_\nu C_\ell^{\rm \phi \times CIB(\nu)} \right) - \alpha_{\rm corr} C_\ell^{\rm \phi \times CIB(857)}.
\end{equation}

In Fig.~\ref{cibres}, we present the obtained residuals for the set of weights from \citet{hil13}. The CIB-$\phi$ residual contributes with $(20 \pm 10)$\% of the tSZ-$\phi$ signal for $\ell$ ranging from 500 to 2000.\\
This level of contamination is dependent on the CIB luminosity function (see Sect.~\ref{2halos}). The uncertainty level was estimated by propagating the CIB luminosity function uncertainties \citep[see][for details on estimating the CIB luminosity function parameters]{PlanckCIB}. This uncertainty level is dominated by $\delta_{\rm cib}$ and $\epsilon_{\rm cib}$, which set the amount of correlation between the CIB maps observed at different frequencies. Other parameters, such as $b_{\rm lin}$ or CIB SED parameters, have a weaker effect on the CIB-$\phi$ residual estimation.

\section{Conclusion and discussion}
\label{seccon}
We have presented a modelling of the tSZ-$\phi$-CIB cross-correlations.
Based on this modelling, we predicted the expected CIB-$\phi$ contamination in the tSZ-$\phi$ power-spectra deduced from a Compton parameter map built by linearly combining {\it Planck} channels from 100 to 857 GHz.\\
We demonstrated that the expected level of contamination from the CIB-$\phi$ is at the same level as the tSZ-$\phi$ signal itself.\\
We also demonstrated that the CIB-$\phi$ contamination level is highly dependent on the set of weights used for constructing of the Compton parameter map. To do this, we used realistic values deduced from the sky, for the weights.\\
We tested a cleaning method for this bias and showed that the level of residual bias can reach about 20\% of the tSZ-$\phi$ spectrum for the used set of weights.
Consequently, we stress that an tSZ-$\phi$ analysis that is based on a multi-frequency Compton parameter map may present a high level of bias. A careful CIB-$\phi$ analysis has to be performed simultaneously to a tSZ-$\phi$ analysis to avoid high bias in the final results.\\
\citet{hil13} constrained $\sigma_8 (\Omega_m/0.282)^{0.26} = 0.824 \pm 0.029$ and found that the tSZ-$\phi$ power-spectrum scales as $\sigma_8^{6.1}$. Thus, a bias of $(20 \pm 10)$\% on the the tSZ-$\phi$ amplitude produces a bias of $(3.3 \pm 1.7)$\% on the $\sigma_8$ amplitude. As a consequence, the CIB-$\phi$ induced residual bias in this measurement contributes about 1 $\sigma$ of the uncertainty. Such a bias may explain the difference between the tSZ-$\phi$ cross-correlation and cosmological parameter estimation made with a tSZ spectrum. The latter favours lower values for $\sigma_8$, with $\sigma_8 (\Omega_m/0.28)^{0.395} = 0.784 \pm 0.016$.

\section*{Acknowledgment}
\thanks{
We are grateful to N.Aghanim and M.Douspis for usefull discutions.\\
  We acknowledge the support of the French \emph{Agence
    Nationale de la Recherche} under grant ANR-11-BD56-015.\\ 
    }

\bibliographystyle{aa}
\bibliography{szlens_conta_cib}

\end{document}